# Estimates of the Maximum Lasing Frequency and of the Gain in FEL with Parabolic Potential

K.S. Badikyan[*], D.K. Hovhannisyan

National University of Architecture and Construction, Yerevan, Armenia

[*] badikyan.kar@gmail.com

## Abstract

The resonance frequency of the system is found and the linear gain is derived for the odd harmonics of this frequency. Averaging of the gain is carried out over the initial distribution of electrons in a transverse cross section of the beam. Estimates are obtained of the maximum lasing frequency and of the gain at this frequency

## 1. Introduction

The quantum theory of amplification in relativistic strophotron is developed in [1]. The equations for the amplitudes of transition probabilities are found. The potential through is used in strophotrons [2-20] in contrast to usual FELs [21-45] where magnetic undulators are used. The classical theory of amplification in RS is developed in [46].

We shall analyze the dependence of the resonance frequency on the initial transverse coordinate of an electron $\omega_{res}(x_0)$. We shall show that this dependence results in a strong inhomogeneous broadening and overlap of the gain profiles at different harmonics of the fundamental resonance frequency. We shall assume these conditions in determining the gain averaged over $x_0$, i.e., over the distribution of electrons in a transverse cross section of the beam. We shall show how the average gain retains its resonance structure, but resonance amplification includes contributions of only a small fraction of the beam electrons that enter the system close to the axis (small values of $x_0$). We shall conclude with numerical estimates of the maximum attainable lasing frequency and of the corresponding gain.

A geometry of the fields similar to that considered in the present study applies also to the case of emission of radiation as a result of channeling of electrons and positrons in crystals

[17,20,23]. In contrast to the conventional channeling, the interaction of electrons with an external macroscopic field (which can be described as macroscopic channeling) is much simpler because there are no such problems as the absorption of electromagnetic waves in a crystal, possible limits on the channeling length, etc.

In the present article the emitted energy is calculated in RS using quantum mechanical approach. The averaging of gain is given and its maximal value is found.

## 2. Resonance frequency and output energy

Equation (8) of [1] gives

$$\omega \approx \omega_{res}^{(s)} = 2(2s+1)\gamma^2\Omega/(1+l_0\Omega\gamma^2/\varepsilon) \equiv (2s+1)\omega_{res}, \quad (1)$$

the resonance frequency of the system $\omega_{res}$ expressed in terms of the number of an oscillatory level $l_0$ filled most effectively at the initial moment in time t = 0.

The value of $l_0$ can be expressed in terms of the initial parameters of the electron beam. A correct determination of $l_0$ requires the knowledge of the initial transverse wave function of an electron in the form of a packet localized at some point $x_0$. Expanding this function in terms of $\varphi_l(x)$ of Eq. (2) of [1], we find that the squares of the module of the coefficients in the expansion have a narrow maximum at

$$l \approx l_0 \equiv p_\perp^2/2\varepsilon + x_0^2\Omega\varepsilon/2 = (\varepsilon/2\Omega)(\alpha^2 + x_0^2\Omega^2) = \varepsilon\Omega a^2/2, \quad (2)$$

where $a(x_0) = (x_0^2 + \alpha^2/\Omega^2)^{1/2}$ is the classical amplitude of transverse oscillations of an electron in a strophotron with an initial coordinate $x_0$ and the angle of entry $\alpha$ into the field.

Substitution of $l_0$ from Eq. (2) into Eq. (1) gives

$$\omega_{res} = 2\lambda^2\Omega/[1+(\gamma\Omega a)^2/2]. \quad (3)$$

This expression for the resonance frequency has been obtained earlier in the theory of spontaneous emission during channeling of particles in a crystal. If $\gamma\Omega a \ll 1$, then the frequency $\omega_{res}$ of Eq. (3) is identical with the result obtained in Ref. [2]: $\omega_{res} = 2\gamma^2\Omega$.

We shall now calculate the energy $\Delta E$ emitted by an electron at a frequency $\omega$ in the direction of the OZ axis and governed by the expression

$$\Delta E = -\omega \sum_n n|a_n|^2. \quad (4)$$

According to Ref. [3], the simplest method of finding ΔE *in an* approximation which is linear in $E_0^2$ involves solution of the system (9) of [1] by perturbation theory in terms of $\tilde{E}_{int}$, which gives

$$\Delta E = \frac{(eE_0)^2 t^3 \omega}{64\varepsilon}(\Omega a)^2 \left[\frac{1}{\gamma^2} + \frac{(\Omega a)^2}{4}\right] F_s^2 \frac{d}{du_s}\left(\frac{\sin^2 u_s}{u_s^2}\right), \qquad (5)$$

where

$$u_s = \frac{-(\Delta_s E_{anh} t)}{\omega} = \frac{t}{4}\left[\omega\left(\frac{1}{\gamma^2} + \frac{(a\,\Omega)^2}{2}\right) - 2(2s+1)\Omega\right],$$

$$\tilde{E}_{int} = (-1)^s E_{int} F_s(z); \quad F_s(z) = J_s(z) - J_{s+1}(z), \qquad (6)$$

$$\Delta_s = (\omega/2E_{anh})\left[(2s+1)\Omega - 1/2\omega\left(1/\gamma^2 + l_0\Omega/\varepsilon\right)\right],$$

$$E_{anh} = (\omega/2\varepsilon)\left[\omega/\gamma^2 - (2s+1)\Omega + 3l_0\Omega\omega/4\varepsilon\right].$$

The quantity $\Delta E_s$ represents the energy emitted by an electron at a frequency $\omega \approx \omega_{res}^s = (2s+1)\omega_{res}$. For fixed values of α and $x_0$ the widths of the gain profiles associated with the finite interaction time between an electron and the system are $\delta\omega \approx \omega_{res}/N_{osc}$, where $N_{osc} = \Omega t/2\pi$ is the number of transverse oscillations of an electron in the transit time *t* across the system. The separation between neighboring gain profiles is $\Delta\omega = 2\omega_{res} \gg \delta\omega$, i.e., in the case of a single electron the lines do not overlap if $N_{osc} = \Omega t/2\pi \gg 1$. However, in view of the dependence $\omega_{res}(x_0)$ of Eq. (3), different electrons (corresponding to different values of $x_0$ have different resonance frequencies $\omega_{res}$ and, consequently, different positions of the resonance amplification lines.

The energy emitted by a beam at a frequency ω can be found by summing Δ $\Delta E_s$ over *s* and averaging the total emitted energy

$$\Delta E = \sum_s \Delta E_s \qquad (7)$$

with respect to $x_0$. This procedure and the corresponding conditions are described in the next section, where the average gain is found. However, we shall first consider the problem of the similarity and difference between free-electron lasers with an undulator and a parabolic trough (strophotron).

It is known that a free-electron laser with a planar linearly polarized undulator can amplify a wave traveling along the axis at a frequency ω close to the odd harmonics of the fundamental resonance frequency. The formulas (3), (5), and (6) governing the energy emitted by a single electron can be reduced to the form similar to the corresponding formulas for a free-electron laser with an undulator [3-5] if we introduce a parameter $K_{str} = \gamma \Omega a / c$, which replaces the usual parameter for an undulator $K_{und} = eB_0 \lambda_0 / mc^2$ ($B_0$ and $\lambda_0$ are the undulator intensity and period). The major differences between these two systems are observed when we consider a beam as a whole rather than one electron. If $K_{und}$ is independent of the initial conditions, then in the system under discussion we have $K_{str} = K_{str}(x_0, \alpha)$. For this reason the averaging over $x_0$ is not as trivial as in the case of an undulator.

The relative efficiency of stimulated emission of the (2s + 1)th harmonic by a single electron is governed by the factor $F_s$ of Eq. (6) in Eq. (5). If $\gamma \Omega a << 1$, then $\omega_{res} = 2\gamma^2 \Omega$ and the factor is $z[\omega \approx (2s+1)\omega_{res}] \approx (1/4)(2s+1)(\gamma \Omega a)^2 << s\zeta$. Since the argument of the Bessel functions in Eq. (6) is small compared with the index, it follows that amplification of high harmonics is then impossible ($\Delta E_s$ falls rapidly on increase in s).

Effective amplification at higher harmonics requires that the parameter $K_{str}$ should be large: $K_{str} >> 1$, which we shall henceforth assume to be correct. If $K_{str} >> 1$, then the resonance frequency $\omega_{res}$ of Eq. (3) differs considerably from $2\gamma^2 \Omega$ ($\omega_{res} << 2\gamma^2 \Omega$). However, the lasing frequencies $\omega_{res}^s = (2s+1)\omega_{res}$ need not be small compared with $2\gamma^2 \Omega$ because the number $s$ is fairly large. The possibility of increasing $s$ is governed, as usual, by the condition of a moderately strong fall of the average gain on increase in *s*.

### 3. Averaging over the distribution of electrons in a beam. Gain

We shall consider a specific distribution of electrons in a beam along the initial transverse coordinate described by the Gaussian function

$$f(x_0) = \frac{1}{\sqrt{\pi}\tilde{d}_e} \exp\left[-\frac{x_0^2}{\tilde{d}_e^2}\right], \quad (8)$$

where $\tilde{d}_e = d_e / (2\sqrt{\ln 2})$ ($d_e$ is the diameter of the electron beam satisfying the condition $d_e < 2d$).

We shall first estimate the scale of inhomogeneous broadening of the lines due to the scatter of $x_0$ and assume that the condition $d_e \Omega / \alpha << 1$ is satisfied.

The change in $\omega_{res}$ of Eq. (3) on variation of $x_0$ from 0 to $d_e/2 \approx d_e$ is equal to $\Delta\omega_{res} \approx (1/16)d_e^2\Omega\omega_{res}^2$. The corresponding shift of the emission line representing the (2s+1)th harmonic is $\omega_s = (2s+1)\omega_{res} \approx \frac{1}{16}(2s+1)d_e^2\Omega\omega_{res}^2$.

This shift exceeds the homogeneous line width $\delta\omega \approx \omega_{res}/N_{osc}$, if the parameter of the distribution (8) satisfies the inequality

$$d_e^2 > \Delta x_0^2 \equiv 16/N_{osc}\Omega\omega. \tag{9}$$

If

$$d_e^2 > \Delta x_1^2 \equiv 32/\Omega\omega,$$

then $\Delta\omega_{res}$ exceeds also dissipation between the neighboring emission lines $\Delta\omega \approx 2\omega_{res}$ and these lines overlap due to inhomogeneous broadening.

Clearly, if the condition $d_e < \Delta x_0$ is obeyed, an inhomogeneous broadening plays no significant role and averaging over $x_0$ does not alter $\Delta E$ or, consequently, the gain. Therefore, the gain found in Ref. [5] is correct if $K_{str} \ll 1$ and $d < \Delta x_0$.

For sufficiently large values of $\Omega$ and $t$, necessary to ensure acceptable values of the lasing frequency $\omega$ and of the gain, the parameter $\Delta x_0$ is very small so that the condition (9) is always satisfied by real beams.

We shall now carry out averaging of the total output energy $\Delta E$ of Eq. (7) integrating term by term this sum with respect to $x_0$ allowing for the distribution function (8). In each term of the sum (7) the main contribution to the integrals with respect to $x_0$ is made by small ($\sim \Delta x_0$) regions near the points $x_0^{(s)}$ such that $u_s(x_0^{(s)}) = 0$:

$$(x_0^{(s)})^2 = \frac{4}{\Omega\omega}(2s+1) - \frac{\alpha^2}{\Omega^2} - \frac{2}{\gamma^2\Omega^2} \equiv \frac{8}{\Omega\omega}(s - s_{min}) \geq 0, \tag{10}$$

where

$$s_{min} \equiv \frac{\alpha^2\omega}{8\Omega} + \frac{\omega}{4\gamma^2\Omega} - \frac{1}{2}. \tag{11}$$

The argument of the Bessel functions $z(x_0)$ at the point $x_0 = x_0^{(s)}$

$$z(x_0 = x_0^{(s)}) \equiv z_s = s + \frac{1}{2} - \frac{\omega}{4\gamma^2\Omega}. \tag{12}$$

In view of the need to consider the possibility of amplification of higher harmonics [$s \gg (\alpha\gamma)^3$], when $z_s \leq s$ and $z_s, s \gg 1$, we shall express the Bessel functions in terms of the MacDonald functions [47]; then, the relevant factor is

$$F_s \approx \frac{2(s-z_s)}{\pi\sqrt{3}s} K_{2/3}\left[\frac{2(s-z_s)^{3/2}}{3\sqrt{s}}\right] \approx \frac{\omega K_{2/3}\left[\frac{\omega}{\omega_{max}}\sqrt{\frac{s_{min}}{s}}\right]}{2\sqrt{3}\pi\gamma^2\Omega s}, \tag{13}$$

where

$$\omega_{max} = 3\gamma^3 \alpha \Omega. \tag{14}$$

The range of values of $s$ that make the main contribution to the average total energy ($\langle\Delta E\rangle$), emitted at a given frequency $\omega$ is found from the condition $s > s_{min}$. At a fixed frequency $\omega$ the term in Eq. (7) with the number $s = s_{min}$ becomes the dominant one after averaging with respect to $x_0$. This means that the resonance radiation appears mainly because of electrons which enter a strophotron near the axis ($x_0 \approx 0$) and are characterized by the coordinate scatter $\Delta x_0$. Such radiation is emitted at frequencies $\omega \approx \omega_{res}^{(s)}(x_0 = 0) = (2s+1)\omega_{res}(x_0 = 0) \approx (8s\Omega)/\alpha^2$.

Averaging of all the other terms with $s > s_{min}$ gives rise to a nonresonance background, the intensity of which is less than the resonance value because of the factor $N_{osc}^{-3/2} \ll 1$.

The final result of averaging the sum (7) with respect to $x_0$ can be represented in the form

$$\langle\Delta E\rangle = \frac{(eE_0)^2 t^{5/3} \left[\omega_{res}^{(s)}(x_0 = 0)\right]^{1/2}}{24\sqrt{2}\pi^{5/2}\tilde{d}_e\varepsilon\gamma^4\Omega} K_{2/3}^2\left[\frac{\omega}{\omega_{max}}\right] F_{res}(\omega), \tag{15}$$

where

$$F_{res}(\omega) = \int_{-\infty}^{\infty} \frac{d\xi}{\xi} \frac{d}{d\xi} \frac{\sin^2\left[\xi^2 + \frac{\alpha^2 t}{8}\left(\omega - \omega_{res}^{(s)}(x_0 = 0)\right)\right]}{\left[\xi^2 + \frac{\alpha^2 t}{8}\left(\omega - \omega_{res}^{(s)}(x_0 = 0)\right)\right]^2} \tag{16}$$

is the form factor of the resonance curve of width $\Delta\omega \sim \frac{8}{\alpha^2 t} = \frac{2\omega_{res}(x_0 = 0)}{\Omega t} \ll 2\omega_{res}(x_0 = 0)$ and the value of the maximum is $\left[F_{res}(\omega)\right]_{max} \simeq 0.5$.

Finally, the gain experienced by an external wave of frequency $\omega$ is given by the expression

$$G = \eta\frac{8\pi N_e}{E_0^2}\langle\Delta E\rangle = \eta\frac{(2\ln 2)^{1/2}}{\sqrt{2}\pi^{3/2}} \frac{N_e r_0 L^{5/3}\alpha^{1/2}}{d_e\gamma^{7/2}\Omega^{1/2}} \left(\frac{\omega}{\omega_{max}}\right)^{1/2} K_{2/3}^2\left[\frac{\omega}{\omega_{max}}\right] F_{res}(\omega), \tag{17}$$

where $N_e$ is the density of electrons in a beam; $r_0 = e^2/mc^2$ is the classical radius of an electron; $\eta = d_e/2a(x_0 = 0) = d_e\Omega/2\alpha$ is a factor which allows for the overlap of the electron beam and the wave being amplified.

The dependence of the gain $G$ of Eq. (17) on the frequency $\omega$ exhibits a steep fall at $\omega > \omega_{max}$, when the argument of the MacDonald function becomes large. Therefore, $\omega_{max}$ of Eq. (14) is the maximum frequency in a free-electron laser of the strophotron type and right up to this frequency the gain can be significant.

## 4. Conclusions

The gain of an external wave in the strophotron is given by a superposition of contributions from amplification at different (odd) harmonics of the main resonance frequency $\omega_{res}(x_0)$. The main resonance frequency is shown to depend on the initial conditions of the electron, and in particular on its initial transversal coordinate $x_0$. This dependence $\omega_{res}(x_0)$ is shown to give rise to a very strong inhomogeneous broadening of the spectral lines. The broadening can become large enough for the spectral lines to overlap with each other. The gain is averaged over $x_0$. In the averaged gain the resonance peaks are shown to be much higher than the nonresonant background. A physical nature of these resonances remaining after averaging is discussed. The maximum achievable averaged gain and frequency of the strophotron FEL are estimated.

Assuming that $\omega \approx \omega_{max}$, $\alpha \approx d_e\Omega$, ($\eta \approx 0.5$), ($K_{2/3}(1) \approx 0.5$, and $[F_{res}(\omega)]_{max} \approx 0.5$, we can write Eqs. (3) and (17) in the form

$$\omega = 5.3 \times 10^7 \gamma^2 d_e g, \tag{18}$$

$$G = 7.6 \times 10^{-3} N_e r_0 L^{5/2} / d_e^{1/2} \gamma^{7/2} = 2.1 \times 10^6 J_{max} r_0 L^{5/2} / d_e^{5/2} \gamma^{7/2} \tag{19}$$

where $g$ is the gradient of the field in a parabolic potential trough on the axis of the system $OZ$ (expressed in gauss per centimeter in the case of magnetic quadrupole lenses); $J_{max}$ is the total maximum current in the beam (in amperes). It is interesting to note that the gain $G$ of Eq. (19) at the maximum frequency $\omega = \omega_{max}$ is independent of $g$.

Adopting in these estimates the value $d_e = 1 cm$, $J_{max} = 100A$, $\gamma = 10$ A, $L = 2$ m, and $g = 10$ kG/cm, we find that

$$\omega = 5.3 \times 10^{13} \sec^{-1}(\lambda = 36\mu), \quad G \approx 1\%.. \tag{20}$$

These estimates show that in the case of the above parameters of the beam and other components of the system, it is in principle possible to construct a strophotron free-electron laser operating in the infrared range.

We shall now note the singularities which distinguish a strophotron free-electron laser from a conventional undulator laser.

1. The undulator parameter $K_{str}$ depends only on the initial conditions and not on the relativistic factor $\gamma$; it can be numerically large. In the above example, we have $K_{str}(x_0 = 0) = \alpha\gamma \approx 7.7$. The high value of $K_{str}$ shows that the maximum serial number of the harmonic in which more or less effective amplification is still possible is high: $s_{max} = \omega_{max}/2\omega_{res}(x_0 = 0) \approx 0.3[(1/2)K_{str}(x_0 = 0)]^3 \approx 168$.

   The gain profile extends from $\omega_{res}(x_0 = 0)$ to $2s_{max}\omega_{res}(x_0 \approx 0)$ and it consists of a large number of equidistant narrow gain lines.

2. The frequency of a wave amplified in a strophotron depends strongly on the angle at which an electron enters a system. For a fixed value of $s$, we have $\omega \propto 1/\alpha^2$, whereas at $s = s_{max}$, we find that $\omega \propto \alpha$. Hence, it follows that a strophotron provides an opportunity for continuous tuning of the emission frequency because of a change in the entry angle $\alpha$.

   Clearly, there is a margin for increasing the gain. In particular, preliminary shaping of an electron beam so that the angle of entry of electrons into a strophotron depends on the initial transverse coordinate $x_0$ is a promising approach. In this sense the optimal distribution is that for which the quantity $\alpha^2 + x_0^2\Omega^2$ remains constant over the whole beam diameter. This can be achieved by, for example, use of a focusing device based on electron-optical lenses.

   In a system of this kind there is no strong inhomogeneous broadening because of the distribution of electrons over $x_0$, so that the gain of Eq. (19) increases by a factor $(d_e/\Delta x_0)^2$, i.e., by the two orders of magnitude.

   A quantum-mechanical description of the motion of an electron in classical fields is not in conflict with the fact that the final results obtained here do not contain the Planck constant $\hbar$. It is natural to expect the main results of the present study apply also in the classical approach.